\documentclass[prl,10pt,twocolumn,showpacs,superscriptaddress]{revtex4}
\usepackage{graphicx}
\usepackage{amsmath,amssymb,amsfonts}

 \newcommand{\nc}{\newcommand}
 \nc{\mb}[1]{\makebox[#1]{}}
 \nc{\V}{{\rm v}}
 \nc{\W}{{\scriptscriptstyle W}}
 \nc{\X}{{\scriptscriptstyle X}}
 \nc{\CSV}{{\scriptscriptstyle CSV}}
 \nc{\be}{\begin{equation}}
 \nc{\ee}{\end{equation}}
 \nc{\bea}{\begin{eqnarray}}
 \nc{\eea}{\end{eqnarray}}
 \nc{\ra}{\rightarrow}
 \nc{\alS}{{\alpha_{\scriptscriptstyle S}}}
 \nc{\aSpi}{{\frac{\alS}{2\pi}}}
 \nc{\api}{{\frac{\alpha}{2\pi}}}
 \nc{\dwtilm}{{\delta \widetilde{m}}}
 \nc{\ppg}{\pi^+\pi^-\gamma}
 \nc{\nubar}{{\overline{\nu}}} 
 \nc{\nuN}{{\nu N_0}}
 \nc{\nubN}{{\overline{\nu} N_0}}
 \nc{\snuNC}{{\langle \sigma^{\nuN}_{\NC}\rangle }}
 \nc{\snubNC}{{\langle \sigma^{\nubN}_{\NC}\rangle }}
 \nc{\snuCC}{{\langle \sigma^{\nuN}_{\CC}\rangle }}
 \nc{\snubCC}{{\langle \sigma^{\nubN}_{\CC}\rangle }}
 \nc{\snNC}{{\langle \sigma^{\nu p}_{\NC}\rangle }}
 \nc{\snbNC}{{\langle \sigma^{\nubar p}_{\NC}\rangle }}
 \nc{\snCC}{{\langle \sigma^{\nu p}_{\CC}\rangle }}
 \nc{\snbCC}{{\langle \sigma^{\nubar p}_{\CC}\rangle }}
 \nc{\Rnu}{{R^{\nu}}}
 \nc{\Rnub}{{R^{\overline{\nu}}}}
 \nc{\sintW}{{\sin^2 \theta_{\W} }}
 \nc{\vp}{{\bf p}}
 \nc{\uv}{{u_{\rm v}}}
 \nc{\dv}{{d_{\rm v}}}
 \nc{\ubar}{{\overline{u}}} 
 \nc{\dbar}{{\overline{d}}} 
 \nc{\sbar}{{\overline{s}}} 
 \nc{\cbar}{{\overline{c}}} 
 \nc{\Ubar}{{\overline{U}}} 
 \nc{\Dbar}{{\overline{D}}} 
 \nc{\Sbar}{{\overline{S}}} 
 \nc{\Qbar}{{\overline{Q}}} 
 \nc{\FbWp}{{\overline{F}_2^{Wp}}}
 \nc{\FbWD}{{\overline{F}_2^{WD}}}
 \nc{\rz}{{1\over \rho_0^2}}
 \nc{\gLu} {{g_L^u}}
 \nc{\gRu} {{g_R^u}}
 \nc{\gLd} {{g_L^d}}
 \nc{\gRd} {{g_R^d}}
 \nc{\Delu} {{\Delta u^2}}
 \nc{\Deld} {{\Delta d^2}}
 \nc{\Rnp} {{R^{\nu}_p}}
 \nc{\Rnbp} {{R^{\nubar}_p}}
 \nc{\Pcs}{{P_{CS}}}
 \def\CC{{\scriptscriptstyle CC}}

 \def\NC{{\scriptscriptstyle NC}}

 \def\IE{{\it i.e.,\ }}
 
 \def\EA{{\it et al.\ }}

\begin{document}


\author{T.J.~Hobbs}
\affiliation{Department of Physics and Center for Exploration of Energy and 
Matter, Indiana University, Bloomington, IN 47405, USA}

\author{J.~T.~Londergan}
\affiliation{Department of Physics and Center for Exploration of Energy and 
Matter, Indiana University, Bloomington, IN 47405, USA}

\author{D.P.~Murdock}
\affiliation{Department of Physics, Tennessee Technological University, 
Cookeville, TN 38505, USA}

\author{A.~W.~Thomas}
\affiliation{CSSM, School of Chemistry and Physics, University of Adelaide, 
Adelaide, South Australia 5005, Australia}

\title{Testing Partonic Charge Symmetry at a High-Energy Electron Collider}

\begin{abstract}
We examine the possibility that one could measure partonic charge symmetry 
violation (CSV) by comparing neutrino or antineutrino production through 
charged-current reactions induced by electrons or positrons at a possible 
electron collider at the LHC. We calculate the magnitude of CSV that might 
be expected at such a facility. We show that this is likely to be a 
several percent effect, substantially larger than the typical CSV effects  
expected for partonic reactions. 
\end{abstract}

\maketitle


Charge symmetry is a very specific operation involving isospin, which leads  
to the interchange of protons and neutrons, or equivalently the 
interchange  of up and down quarks. The charge symmetry operator $\Pcs$ 
corresponds to a rotation of $180^{\circ}$ about the $2$ axis in isospin 
space, such that 
\bea
\Pcs &=& e^{i\pi T_2} \ , \nonumber \\ 
\Pcs |u\rangle &=& -|d\rangle; \hspace{0.6cm} \Pcs |d\rangle = |u\rangle .
\label{eq:PCSdef}
\eea 
It is of particular importance because at low energies, where it has been 
studied extensively, charge symmetry is a far better symmetry than isospin 
in general, typically being respected to better than 
1\%~\cite{Henley:1979,Miller:1990iz}. It is therefore natural to assume 
that charge symmetry is also valid at the partonic level and, indeed, almost 
all analyses of parton distribution functions (PDFs) assume charge symmetry, 
whether the assumption is stated or not. The importance of charge symmetry 
violation in PDFs within the context of tests of the Standard Model has 
recently been of considerable interest~\cite{Bentz:2009yy,Londergan:2003ij}.

To date, no 
violation of charge symmetry has been observed at the partonic level, 
although the one global analysis that did allow for CSV did find a preferred 
solution with a non-zero effect -- albeit with very 
large errors~\cite{MRST03}.
The current upper limits are consistent with the validity of partonic charge 
symmetry in the range 5-10\%~\cite{Londergan:2009kj}. Theoretical models  
generally produce estimates of  charge symmetry violation (CSV) in PDFs which 
for many observables give effects at roughly 
the 1\% level~\cite{Londergan:2009kj,Lo98a}. This presents a significant challenge 
for experimentalists, first to observe effects of this magnitude and 
then to isolate the signal from competing effects of similar size. 

A new facility has recently been proposed that would collide electrons 
or positrons from an electron accelerator with protons or deuterons from 
the LHC~\cite{LHeC}. In this paper we will show that such a 
facility (given the name LHeC) has the potential to produce charge symmetry 
violating effects which are considerably larger than those expected with 
other facilities. We will review the effect in question, show the results 
of theoretical calculations for the proposed CSV effects, and discuss why 
they ought to be expected to be relatively large at energies 
accessible to an electron-ion collider.     

\begin{figure}[ht]
\includegraphics[width=3.0in]{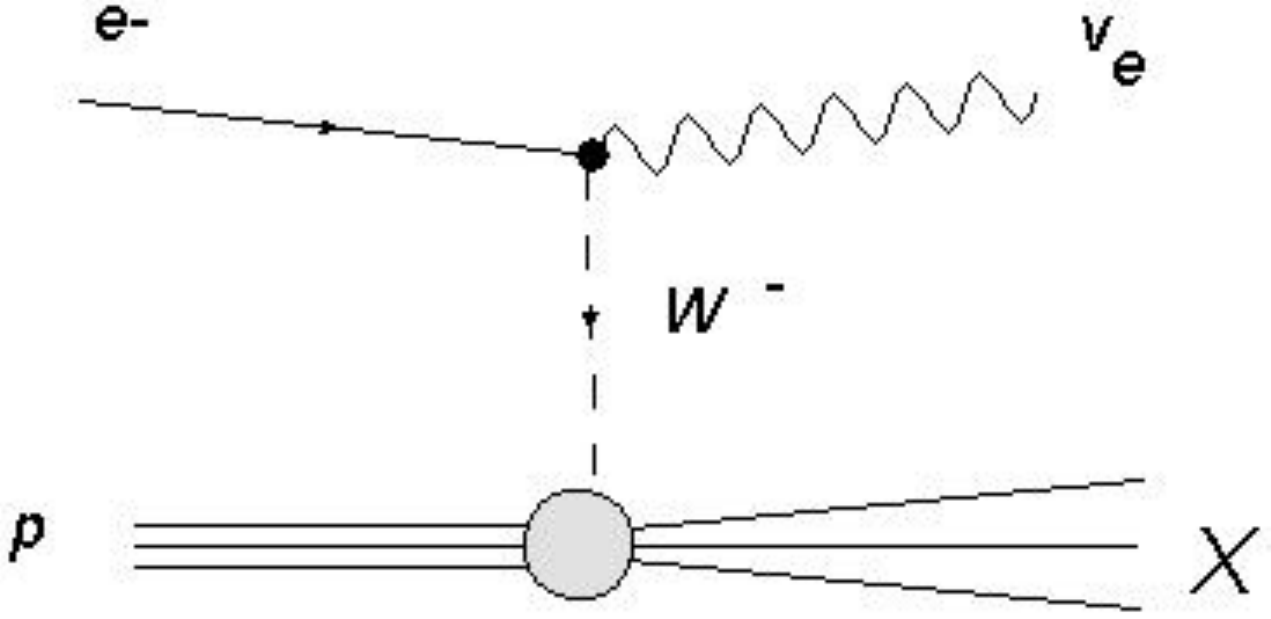}
\caption{Schematic picture of charged-current neutrino production in DIS 
induced by an electron on a proton.} 
\label{Fig:fig1}
\end{figure}

The reactions of interest are the charged current (CC) cross sections for 
electron and positron deep inelastic scattering at energies in the range 50-100 GeV 
on protons and deuterons at LHC energies, \IE several TeV. These are important 
because they directly and unambiguously probe the flavor structure of the 
proton PDFs in the valence region. Consider the 
deep inelastic reaction $(e^-,\nu_e)$. A high-energy electron incident 
on a proton produces a neutrino. The process results from a $W^-$ which 
is absorbed on quarks from the proton, as shown schematically in 
Fig.~\ref{Fig:fig1}; the final hadronic state is not observed. The signature 
for this process is disappearance of the electron, together with very large 
deposition of energy in the hadronic sector. 

The $F_2$ structure function for the CC reaction on a proton has the form   
\be
F_2^{W^- p}(x) = 2x[ u(x) + c(x) + \dbar(x) + \sbar(x)] .  
\label{eq:F2ep}
\ee
These reactions will occur at extremely high energies 
and very large $Q^2$. Therefore, any corrections to the $F_2$ structure 
functions in Eqs.~(\ref{eq:F2ep}) and (\ref{eq:F2ePp}) 
arising from quark mixing matrices, quark masses or higher-twist effects 
should be completely negligible. 

We can also consider the corresponding reaction for positrons on protons, 
$(e^+,\nubar_e)$. This reaction involves the absorption of a $W^+$ on the 
proton, with the resulting $F_2$ structure function  
\be
F_2^{W^+ p}(x) = 2x[ \ubar(x) + \bar{c}(x) + d(x) + s(x)] .  
\label{eq:F2ePp}
\ee
We can straightforwardly calculate the $F_2$ structure functions (per nucleon) 
on the deuteron,  
\bea
F_2^{W^- D}(x) &=& x[ u^+(x) + d^+(x) + 2c(x) \nonumber \\ &+& 
2\sbar(x)- \delta d(x) - \delta\ubar(x)]\ ; \nonumber \\  
F_2^{W^+ D}(x) &=& x[ u^+(x) + d^+(x) + 2\bar{c}(x) \nonumber \\ &+& 2s(x)- 
\delta\dbar(x) - \delta u(x)]\ .  
\label{eq:F2eD}
\eea

In Eq.~(\ref{eq:F2eD}) we introduce combinations of quark parton distribution 
functions (PDFs) that are even or odd under charge conjugation, and the CSV 
PDFs 
\bea
 q^{\pm}(x) &=&  x[q(x) \pm \bar{q}(x)] \ ; \nonumber \\ 
\delta u(x) &=& u^p(x) - d^n(x) \ ; \nonumber \\ 
\delta d(x) &=& d^p(x) - u^n(x) \ .  
\label{qpmdef}
\eea
There are analogous relations to Eq.~(\ref{qpmdef}) for the antiquark CSV 
PDFs. For the remainder of this letter we assume that $\cbar(x) = c(x)$. 
The distributions $q^-(x)$, which involve the differences between quark 
and antiquark PDFs (alternatively, they are the $C$-odd combinations of 
quark distributions), are the \textit{valence} parton distributions for a 
given quark flavor.   

Now define the following quantity, 
\be
R^-(x) \equiv \frac{2(F_2^{W^- D}(x) - F_2^{W^+ D}(x))}{F_2^{W^- p}(x) + 
F_2^{W^+ p}(x)} .
\label{eq:Rmin}
\ee
The quantity $R^-(x)$ is given by the difference in the $F_2$ structure 
functions per nucleon for electron-deuteron and positron-deuteron CC 
reactions, divided by the average $F_2$ structure function for CC 
reactions on protons initiated by electrons and by positrons.  

Using Eqs.~(\ref{eq:F2ep}), (\ref{eq:F2ePp}) and (\ref{eq:F2eD}) we can 
straightforwardly show that the quantity $R^-(x)$ in (\ref{eq:Rmin}) has the 
form 
\be
 R^-(x) = \frac{x[ -2s^-(x) + 
  \delta u^-(x) - \delta d^-(x)]}{x[u^+(x) + d^+(x) + s^+(x) + 2c(x)]} \ .
\label{eq:F2rat}
\ee 
Thus $R^-(x)$ is proportional to the valence quark CSV parton distributions 
plus the strange quark asymmetry (the difference between the strange and 
antistrange PDFs). Insofar as the strange quark asymmetry exists, it should 
be large only at quite small Bjorken $x < 0.1$, while theoretical estimates 
of the valence CSV parton distributions 
\cite{Sather:1991je,Rodionov:1994cg} suggest that for $Q^2 \sim 10$ 
GeV$^2$ they peak at values $x \sim 0.4$.     

\begin{figure}[ht]
\includegraphics[width=2.8in]{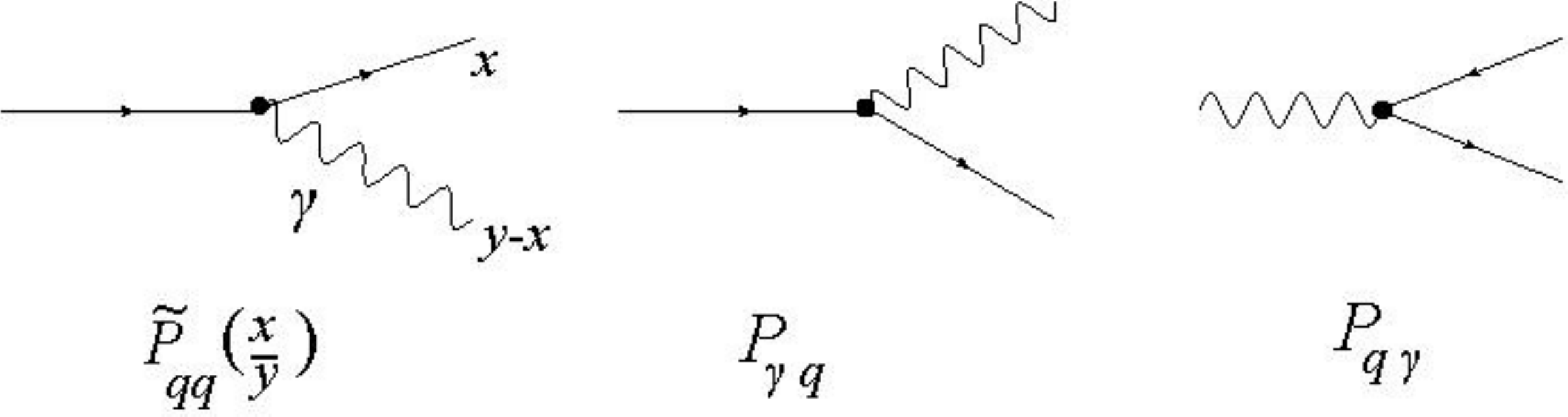}
\caption{Schematic picture of quarks coupling to photons. This gives the 
origin of QED splitting that produces CSV effects in parton 
distribution functions.
 \label{Fig:QEDsplt}}
\end{figure}

Consider a hypothetical collider with 50 GeV electrons or positrons colliding 
with protons and deuterons of energy roughly 7 TeV. This would be 
similar to the possibilities if an electron collider were built at the LHC.  
Consider charged-current reactions at such a facility with $Q^2 = 10^5$ 
GeV$^2$. We know of two different mechanisms for charge symmetry violation 
in parton distribution functions. The first arises from the radiation 
of a photon by a quark. Such contributions are shown schematically in 
Fig.~\ref{Fig:QEDsplt}; they were first calculated by the MRST group 
\cite{Martin:2004dh} and Gluck \EA~\cite{Gluck:2005xh}. These QED corrections are 
analogous to the coupling of gluons to quarks, 
except that photons do not have the self-coupling terms possessed by gluons. 
Inclusion of these `QED splitting' terms will produce charge symmetry 
violation in parton distribution functions because of the electromagnetic (EM) 
coupling due to the different charges of up and down quarks.  

The behavior of parton distributions with increasing $Q^2$ is given by the 
DGLAP evolution equations 
\cite{Dokshitzer:1977sg,Gribov:1972ri,Altarelli:1977zs}. We 
expand the DGLAP evolution equations to lowest order in both the 
strong coupling $\alS$ and the electromagnetic coupling $\alpha$,  
\bea
\frac{\partial q_i(x,\mu^2)}{\partial \ln \mu^2} &=& \aSpi \left[ 
  P_{qq}\otimes q_i + P_{qg}\otimes g \right] \nonumber \\ &+& 
  \api e_i^2 \widetilde{P}_{qq}\otimes q_i \ ; \nonumber \\ 
 \frac{\partial g(x,\mu^2)}{\partial \ln \mu^2} &=& \aSpi \left[ 
  \sum_j P_{gq}\otimes q_j + P_{gg}\otimes g \right] \nonumber \\ &+& 
  \api e_i^2 \widetilde{P}_{qq}\otimes q_i \ ; \nonumber \\ 
\frac{\partial \gamma(x,\mu^2)}{\partial \ln \mu^2} &=& 
  \api \sum_j e_j^2 P_{\gamma q}\otimes q_j \ . 
 \label{eq:DGLAP}
\eea
In Eq.~(\ref{eq:DGLAP}), $q_i(x,\mu^2)$ is the parton distribution 
for a given flavor $i$, $g(x,\mu^2)$ is the gluon distribution and 
$\gamma(x,\mu^2)$ is a ``photon parton distribution'' \cite{Martin:2004dh}. 
In these equations the convolution is defined as 
\be
P\otimes q = \int_x^1 \, \frac{dy}{y} P(y) q(\frac{x}{y}, \mu^2) \ , 
\label{eq:convol}
\ee
and the splitting functions are given by  
\bea
 \widetilde{P}_{qq}(y) &=& \frac{P_{qq}(y)}{C_F}; \hspace{0.3 cm} 
  P_{\gamma q}(y) = \frac{P_{gq}(y)}{C_F}; \nonumber \\    
  P_{q\gamma}(y) &=& \frac{P_{qg}(y)}{T_R}; \hspace{0.3 cm} 
  P_{\gamma \gamma}(y) = \sum_j \frac{-2e_j^2}{3} \delta (1-y) \nonumber \\ 
\label{eq:Pdef}
\eea

We have analogous equations to Eq.~(\ref{eq:DGLAP}) for antiquarks. Taking 
the valence combinations from Eq.~(\ref{qpmdef}), we obtain for up 
and down valence quarks  
\bea
\frac{\partial u^-(x,\mu^2)}{\partial \ln \mu^2} &=&  
 \aSpi P_{qq}\otimes u^- + \frac{2\alpha}{9\pi}\widetilde{P}_{qq}\otimes u^- \ ; 
  \nonumber \\ 
\frac{\partial d^-(x,\mu^2)}{\partial \ln \mu^2} &=&  
 \aSpi P_{qq}\otimes d^- + \frac{\alpha}{18\pi}\widetilde{P}_{qq}\otimes d^- 
  \nonumber \\   
 \label{eq:valevol}
\eea
For the valence CSV parton distributions, since $\delta u^-(x) = u_p^-(x) - 
d_n^-(x)$, from Eq.~(\ref{eq:valevol}) we obtain the evolution equations  
for the valence CSV PDFs, to lowest order in $\alS$ and $\alpha$,       
\bea
\frac{\partial[\delta u^-(x,\mu^2)]}{\partial \ln \mu^2} &\approx& 
  \api (e_u^2 - e_d^2)\widetilde{P}_{qq}\otimes u^- \ ; 
  \nonumber \\ 
 \frac{\partial[\delta d^-(x,\mu^2)]}{\partial \ln \mu^2} &\approx& 
  -  \api (e_u^2 - e_d^2)\widetilde{P}_{qq}\otimes d^- \ . 
 \label{eq:CSVtrunc}
\eea

Eq.~(\ref{eq:valevol}) describes how the valence quarks evolve with 
$Q^2$ and Eq.~(\ref{eq:CSVtrunc}) shows how the valence CSV distributions 
evolve, to lowest order in both $\alS$ and $\alpha$. With increasing $Q^2$,  
partons radiate gluons and photons which carry off momentum. Since the total 
momentum fraction carried by quarks is given by the second moment of the parton 
distributions, as $Q^2$ increases the parton distribution functions will shift 
towards progressively smaller $x$ values. 

Comparison of Eqs.~(\ref{eq:valevol}) and (\ref{eq:CSVtrunc}) shows that the 
radiation from valence quarks will be greater than that from the valence CSV 
distributions. This occurs because to lowest order in $\alS$ and $\alpha$, 
valence quark evolution contains contributions from both gluon and photon radiation, 
whereas the valence CSV distribution has only a term from photon radiation. 
This suggests that with increasing $Q^2$ the valence parton  
distributions would experience a larger shift to low $x$ than will the valence 
CSV distributions. We note that the quantity $R^-(x)$ defined in 
Eq.~(\ref{eq:F2rat}) is proportional to the ratio of valence CSV distributions to 
valence PDFs, at a given $x$ value. If the CSV valence distributions are becoming 
larger relative to the valence PDFs at large $Q^2$, then we expect the quantity 
$R^-(x)$ to grow as $Q^2$ increases; specifically we would expect the ratio to 
increase logarithmically with $Q^2$. 

Eq.~(\ref{eq:CSVtrunc}), the QCD evolution equations for the valence CSV parton 
distributions, have been solved by Gl\"uck \EA~\cite{Gluck:2005xh} and also by the 
MRST group \cite{Martin:2004dh}; the two groups made slightly different 
approximations for the initial conditions. We stress that while the effect of 
photon radiation is clear, it is far less obvious that the boundary conditions 
imposed on the calculations are appropriate. That is, we know of no rigorous 
proof that a low scale, typical of quark models, is the appropriate place to set 
the effect to zero. In the absence of a compelling theoretical derivation it 
is extremely helpful to be able to test the idea experimentally.

The MRST group \cite{Martin:2004dh} did attempt an experimental test of 
this method. Including QED radiation in the DGLAP equations introduces a `photon 
parton distribution' $\gamma(x,Q^2)$, which appears in Eq.~(\ref{eq:DGLAP}). 
The MRST group attempted to identify this quantity in the process $ep \rightarrow 
e\gamma X$ where the final state $e$ and $\gamma$ are produced with equal and 
opposite large transverse momentum. This process has been measured by the ZEUS 
Collaboration in $ep$ collisions at $\sqrt{s}= 300$ and 318 GeV 
\cite{Chekanov:2004wr}. The observed cross sections were in reasonable 
agreement with the MRST calculations but disagreed with calculations done using
the Monte Carlo simulations PYTHIA \cite{Sjostrand:2000wi} and HERWIG 
\cite{Marchesini:1991ch}. It would be useful to have other experimental tests 
of this method for including radiation of photons by partons, and the 
experiment suggested here could provide additional confirmation of this method. 
 
A second source of valence parton CSV arises naturally from the 
mass difference between the $u$ and $d$ quarks and may 
be calculated within light cone quark 
models. In such models the valence quark distribution 
can be expressed as~\cite{Jaffe:1983hp,Signal:1989yc,Schreiber:1991tc} 
\bea
q_{\V}(x, \mu^2) &=& M \sum_X \, |\langle X |\frac{1+ \gamma^0\gamma^3}{2} 
 \psi(0) | N\rangle |^2  \nonumber \\ &\times& \delta (M(1-x) - p_X^+ ) \ . 
\label{eq:qvx}
\eea
Eq.~(\ref{eq:qvx}) denotes the process where a valence quark is removed 
from a nucleon $|N\rangle$, and the result is summed over all final states 
$|X\rangle$. The quantity $p_X^+$ is the energy of the state following removal 
of a valence quark with momentum $k$. The quantity $\mu^2$ represents the 
starting value for the $Q^2$ evolution of the 
parton distribution. Eq.~(\ref{eq:qvx}) is formally exact and provides a natural 
starting point for calculations which preserve the correct support of the 
PDFs. 

Model quark wavefunctions are found to be nearly invariant under the small 
mass changes typical of CSV~\cite{Rodionov:1994cg}, so we concentrate on the 
breaking of partonic charge symmetry associated with energy shifts resulting 
from the $u$ and $d$ quark  
mass differences. In particular, we consider the effect of 
the $n-p$ mass difference $\delta M \equiv 
M_n - M_p = 1.3$ MeV, as well as the difference in diquark masses arising from the 
current quark mass difference between up and down quarks. We define the 
quantity 
\be 
\dwtilm = m_{dd} - m_{uu} \, ,
\label{eq:mtilde}
\ee
for which we have a robust estimate $\dwtilm \sim 4$ MeV \cite{Bickerstaff:1989ch}. 
We determine CSV valence PDFs by calculating the variation of quark model 
parton distributions from Eq.~(\ref{eq:qvx}) 
with respect to these quantities, \IE 
\be 
  \delta q_{\V} \approx \frac{\partial q_{\V}}{\partial (\dwtilm)}\dwtilm 
  + \frac{\partial q_{\V}}{\partial (\delta M)}\delta M \ .
\label{eq:iso}
\ee
From Eq.~(\ref{eq:iso}) the valence charge symmetry violating parton 
distributions are obtained by taking variations with respect to diquark 
and nucleon masses on valence parton distributions from quark models. 
The resulting PDFs account for quark and nucleon mass differences 
that lead to CSV effects. 
  
Valence CSV parton distributions arising from quark mass difference effects 
were calculated by Rodionov \EA~\cite{Rodionov:1994cg}. They used  
bag model wavefunctions, including the effect of quark mass differences on the 
quark wave functions as well as on the di-quark and nucleon masses, using 
Eq.~(\ref{eq:qvx}). The Rodionov calculation preserved the 
correct support and included the effect of transverse momentum in the proton.
As in any quark model calculation, the resulting leading twist PDFs 
are appropriate to a relatively low momentum scale (where most of the momentum 
of the nucleon is carried by valence 
quarks~\cite{Schreiber:1990ij,Schreiber:1991tc}). In order to compare 
with experimental data these PDFs are typically evolved up to $Q^2 = 10$ 
GeV$^2$. We subsequently evolved these parton distributions to the higher $Q^2$ 
values appropriate to an electron-ion collider such as the LHeC. 
 
An alternate theoretical approach was due to Sather \cite{Sather:1991je}, 
who investigated the expression for valence parton CSV distributions 
in a static quark picture. In such models the correct support is no longer 
guaranteed. In addition, Sather neglected transverse quark momentum.  
By applying Eq.~(\ref{eq:iso}) to Eq.~(\ref{eq:qvx}) within this 
approximation scheme, Sather obtained an analytic approximation relating 
valence quark CSV to derivatives of the valence PDFs. The analytic 
approximation of Sather is appropriate only at $Q^2$ values appropriate 
for quark model calculations, \IE~$Q^2 \sim 0.25-0.5$ GeV$^2$.    

We used the Sather prescription, differentiating valence parton distribution 
functions to obtain valence CSV PDFs. For this purpose we used the MRST2001 
parton distributions \cite{Martin:2001es} at the starting scale, $Q_0^2 = 1$ 
GeV$^2$. This is slightly too large a value of $Q^2$ for the validity of 
Sather's analytic approximation, but 
the resulting errors should be small. We then inserted the resulting 
CSV PDFs into the DGLAP evolution equations and evolved to the $Q^2$ 
appropriate for the electron collider experiments. The results were similar 
to those obtained using the CSV distributions of Rodionov \EA. 

Since the CSV effects arising from QED splitting effects and from quark 
mass differences are nearly independent, we have added the two effects 
to produce a ``net'' CSV effect.  

The quark model estimates of valence parton CSV can be compared with a recent 
lattice calculation of valence charge symmetry violating parton distributions 
\cite{Horsley:2010th}. The lattice calculation provides striking confirmation 
of the quark model results 
for parton CSV. The lattice calculation was carried out by considering small 
deviations from the $SU(3)$ flavor-symmetric point where the strange and light 
quarks masses are all equal. In that way they could estimate the effects 
of quark mass differences on parton distributions. The lattice results gave 
estimates for the second moment of the valence CSV distributions 
\bea 
\delta U^+ &=& \int_0^1 x\delta u^+(x)\, dx = -0.0023(6) \ , \nonumber \\    
\delta D^+ &=& \int_0^1 x\delta d^+(x)\, dx = +0.0020(3)\ . 
\label{eq:lattUD}
\eea
The lattice results were obtained at a momentum scale $Q^2 = 4$ GeV$^2$. Note 
that the results are appropriate for the $C$-even combination of quarks rather 
than the desired $C$-odd combination for valence quarks, because the lattice 
calculations are sensitive to the $C$-even combination. By comparison, the 
quark model valence quark calculations at a similar scale obtained 
$\delta U^- = -0.0014$ and 
$\delta D^- = +0.0015$ \cite{Rodionov:1994cg,Londergan:2003pq}. Note that the 
lattice result agrees with the quark model results in both the sign and relative 
magnitude of the second moment of the valence CSV distributions. The lattice 
results are 30-50\% larger than the quark model values. The differences may result 
from the inclusion of singlet contributions in the lattice calculations.   
The lattice results are also in good agreement with the best value 
obtained for valence quark charge symmetry violation in a phenomenological 
global fit to high energy data by the MRST group \cite{MRST03}. However, 
the uncertainties in the lattice calculation are considerably smaller than those 
from the global fit.    

There is one final term that enters into the quantity $R^-(x)$ of 
Eq.~(\ref{eq:F2rat}), namely  
the strange quark momentum asymmetry~\cite{Signal:1987gz,Thomas:2000ny}
\be 
xs^-(x) \equiv x[s(x) - \overline{s}(x)]\ .
\label{eq:sminus}
\ee
Strange (antistrange) parton distributions can be measured through 
opposite-sign dimuon production initiated by neutrinos (antineutrinos). 
A neutrino undergoes a charged-current reaction, producing a $\mu^-$ and 
a $W^+$, which is absorbed on an $s$ quark producing a charm quark. The 
charm quark subsequently undergoes a semileptonic decay producing a 
$\mu^+$ and an $s$ quark. The cross section for this process is 
proportional to the strange quark distribution. The corresponding reaction 
initiated by an antineutrino measures the antistrange PDF. 

Dimuon cross sections have been measured by the CCFR \cite{Bazarko:1994tt} 
and NuTeV \cite{Goncharov:2001qe} experiments. From these reactions one 
can extract the quantity $xs^-(x)$. These analyses have been undertaken 
by five groups: CTEQ \cite{Lai:2007dq}; Mason \EA~\cite{Mason:2007zz}; 
the NNPDF Collaboration \cite{Ball:2009mk}; MSTW \cite{Martin:2009iq}; 
and Alekhin, Kulagin and Petti \cite{Alekhin:2009mb}. We used the results 
of the analysis of Mason \EA~of the NuTeV neutrino reactions 
\cite{Mason:2007zz}. We made an analytic fit to the best-fit result from 
Mason \EA~corresponding to $Q^2 = 16$ GeV$^2$. The fit had the form 
\be 
xs^-(x) = A x^b exp(-cx)(x - 0.004) \ .
\label{eq:sfit}
\ee
The resulting strange quark asymmetry was inserted into the DGLAP evolution 
equation and evolved to high $Q^2$. 

 \begin{figure}[ht]
\includegraphics[width=2.8in]{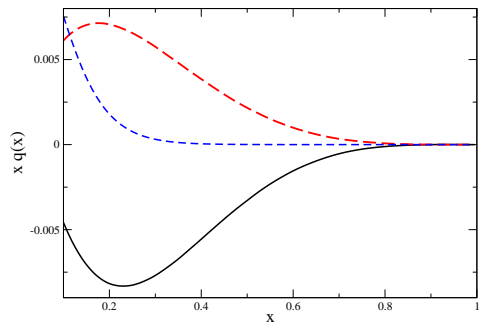}
\caption{[color online] Parton distributions that occur in the numerator 
of Eq.~(\protect\ref{eq:F2rat}). Solid curve: $x\delta u^- (x)$; long-dashed 
curve: $x\delta d^- (x)$; short-dashed curve: $xs^-(x)$. The PDFs have been evolved 
to $Q^2 = 10^5$ GeV$^2$. 
 \label{Fig:numer}}
\end{figure}

The parton distribution functions that occur in the numerator of 
Eq.~(\ref{eq:F2rat}) are plotted in Fig.~\ref{Fig:numer}. The solid curve 
is $x\delta u^- (x)$, the long-dashed curve is $x\delta d^- (x)$ and the 
short-dashed curve is $xs^-(x)$. As one might expect, 
the valence CSV distributions peak at a relatively large 
value $x \sim 0.2$ while the strange quark asymmetry peaks at an extremely 
small $x$ value. Note that due to valence quark normalization, all of 
these quantities must have zero first moment, \IE $\langle q(x) \rangle 
= 0$, where $q = [\delta u^-, \delta d^-, s^-]$. The strange quark 
asymmetry has zero first moment because the proton has no net strangeness; 
the valence CSV distributions must have zero first moment because otherwise 
this would change the total number of valence quarks in the neutron. So 
each of these curves crosses zero at a small value of $x$ (not shown in 
Fig.~\ref{Fig:numer}).

\begin{figure}[ht]
\includegraphics[width=2.8in]{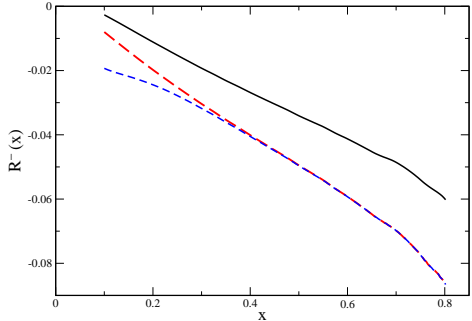}
\caption{[color online] Contributions to the quantity $R^-(x)$ vs.~$x$ from 
Eq.~(\protect\ref{eq:F2rat}), where the PDFs are evolved to $Q^2 = 10^5$ GeV$^2$. 
Solid curve: contribution from QED splitting parton CSV term only; long-dashed 
curve: includes contribution also from quark mass CSV term; short-dashed curve: 
contribution from all terms including strange quark asymmetry.  
 \label{Fig:ratio}}
\end{figure}

Another notable point is that the signs of these quantities are such that (for 
values of $x$ above the crossover point for all of the parton 
distributions)   
all three contributions should add together in the numerator of 
Eq.~(\ref{eq:F2rat}). Fig.~\ref{Fig:ratio} shows the expected value of 
$R^-(x)$ vs.~Bjorken $x$. The solid curve in Fig.~\ref{Fig:ratio} includes 
only the QED splitting contribution to partonic CSV. The long-dashed curve 
includes both QED splitting and quark mass contributions to valence quark 
CSV. The short-dashed curve is the result including all three terms in the 
numerator of Eq.~(\ref{eq:F2rat}), including also the contribution from 
strange quark asymmetry.  
    
We see that for large $x > 0.2$ the strange quark contribution is essentially 
negligible. The predicted values of $R^-(x)$ are large; for $x = 0.6$ the 
calculated ratio is greater than 6\%. This is quite a sizeable result for 
partonic CSV terms, which for most observables yield effects at the 1\% level or 
smaller~\cite{Londergan:2009kj}. This confirms our argument that, while both 
the valence quark and valence CSV distributions shift to lower $x$ values with 
increasing $Q^2$, the CSV distributions experience a smaller shift 
(because to lowest order the CSV valence distributions only radiate 
photons while the valence parton PDFs radiate both gluons and photons), 
and thus 
for a given $x$ the ratio of valence CSV distributions to valence PDFs should 
increase slowly with $Q^2$. Our best theoretical estimate of the ratio 
$R^-(x)$ from 
Eq.~(\ref{eq:F2rat}) at large $x$ values is predicted to be rather large, of 
the order of several percent. For reasonably large values $x > 0.1$, the 
ratio $R^-(x)$ is composed of relatively equal contributions from valence 
parton CSV effects arising from quark mass differences and from QED radiation. 
Thus the quantitative values obtained for the ratio $R^-(x)$ can provide a 
further check on the assumptions made in determining charge symmetry violation 
arising from QED radiation.   

In conclusion, a high energy electron/positron collider whose beams interact 
with deuteron beams from the LHC may produce the most promising observable 
with which to search for partonic charge symmetry violating effects. 

\begin{acknowledgments}
This work was supported in part by the the U.S. National Science Foundation 
grant NSF PHY-0854805 (JTL) as well as by the Australian Research Council 
(through an Australian Laureate Fellowship) and the University of Adelaide 
(AWT).
\end{acknowledgments}

\end{document}